\documentclass[10pt,a4paper,oneside]{article}

\usepackage{float}
\usepackage[ruled,vlined]{algorithm2e}
\usepackage{graphicx}
\usepackage{multirow, graphicx, amsfonts, amsmath}
\usepackage{tikz}
\usetikzlibrary{datavisualization}
\usetikzlibrary{datavisualization.formats.functions}
\usetikzlibrary{arrows,%
	petri,%
	topaths}%

\usepackage{amsmath,amssymb,amsfonts}
\usepackage{algorithmic}
\usepackage{graphicx}
\usepackage{textcomp}
\usepackage{xcolor}
\usepackage{booktabs}
\usepackage[hidelinks]{hyperref}
\usepackage{authblk}

\tikzset{test/.style={font=\footnotesize}}

\providecommand{\keywords}[1]{\textbf{\textit{Index terms---}} #1}

\tikzset{
  basic/.style  = {draw, text width=2cm, drop shadow, font=\sffamily, rectangle},
  root/.style   = {basic, rounded corners=2pt, thin, align=center,
                   fill=green!30},
  level 2/.style = {basic, rounded corners=6pt, thin,align=center, fill=green!60,
                   text width=8em},
  level 3/.style = {basic, thin, align=left, fill=pink!60, text width=6.5em}
}

\begin{document}

\title{A Practical Deep Learning-Based Acoustic Side Channel Attack on Keyboards}
%
%

\author[1]{Joshua Harrison}
\author[2]{Ehsan Toreini}
\author[3]{Maryam Mehrnezhad}
\affil[1]{Durham University, joshua.b.harrison@durham.ac.uk}
\affil[1]{University of Surrey, e.toreini@surrey.ac.uk}
\affil[1]{Royal Holloway University of London, maryam.mehrnezhad@rhul.ac.uk}




\maketitle

\begin{abstract}
With recent developments in deep learning, the ubiquity of microphones and the rise in online services via personal devices, acoustic side channel attacks present a greater threat to keyboards than ever. 
This paper presents a practical implementation of a state-of-the-art deep learning model in order to classify laptop keystrokes, using a smartphone integrated microphone. When trained on keystrokes recorded by a nearby phone, the classifier achieved an accuracy of $95\%$, the highest accuracy seen without the use of a language model. When trained on keystrokes recorded using the video-conferencing software Zoom, an accuracy of $93\%$ was achieved, a new best for the medium. 
Our results prove the practicality of these side channel attacks via off-the-shelf equipment and algorithms. We discuss a series of mitigation methods to protect users against these series of attacks. 

\end{abstract}

\section{Introduction}
Side channel attacks (SCAs) involve the collection and interpretation of signals emitted by a device \cite{24}. Such attacks have been successfully implemented utilising a number of emanation types, such as electromagnetic (EM) waves \cite{31}, power consumption \cite{33}, mobile sensors \cite{mehrnezhad2018stealing,mehrnezhad2016touchsignatures,mehrnezhad2015touchsignatures}, as well as sound \cite{2}. With such a wide range of available mediums, target devices have been similarly varied, with compromised devices including printers \cite{1}, the Enigma machine \cite{30} and even Intel x86 processors \cite{32}. 
It was found in \cite{31} that wireless keyboards produce detectable and readable EM emanations, however there exists a far more prevalent emanation that is both ubiquitous and easier to detect: keystroke sounds \cite{27}. 
The ubiquity of keyboard acoustic emanations makes them not only a readily available attack vector, but also prompts victims to underestimate (and therefore not try to hide) their output. For example, when typing a password, people will regularly hide their screen but will do little to obfuscate their keyboard's sound.
The lack of concern regarding keyboard acoustics could be due to the relatively small body of modern literature. While multiple papers have created models capable of inferring the correct key from test data, these models are often trained and tested on older, thicker, mechanical keyboards with far more pronounced acoustics than modern ones, especially laptops. 

While keyboards have gotten less pronounced over time, the technology with which their acoustics can be accessed and processed has improved dramatically. Examples include advancements in microphone technology, with Voice over Internet Protocol~(VoIP) calls \cite{16} and smartwatches \cite{12} being used to collect keystroke recordings.

Deep Learning (DL) is a subsection of machine learning (ML), in which the model consists of multiple layers of connected neurons. 
Despite being prevalent in the field of computing since the 1960s, DL saw a boom in research in the 2010s benefiting from improvements in graphics processing technology and resulting in huge advances in image recognition \cite{35}, the invention of Generative Adversarial Networks \cite{36} and the invention of transformers \cite{37}. This trend in the performance improvement continues still, with the recent development of the state-of-the-art CoAt Network for image recognition \cite{29}, which combines more traditional convolutional models with transformers. This improvement in DL performance coincides with an increase in access to DL tools. Python packages such as PyTorch \cite{34} provide free and near-universal access to the tools required to run these models on most devices.
With the recent developments in both the performance of (and access to) both microphones and DL models, the feasibility of an acoustic attack on keyboards begins to look likely, as reiterated in recent research \cite{11}. 
While recent papers have explored the viability of ASCAs on laptop keyboards \cite{11,16}, the area remains under-explored considering that laptops make a prime attack vector. Laptops are more transportable than desktop computers and therefore more available in public areas where keyboard acoustics may be overheard, such as libraries, coffee shops and study spaces. 
Moreover, laptops are non-modular, meaning the same model will have the same keyboard and hence similar keyboard emanations. This uniformity within laptops could mean that, should a popular laptop prove susceptible to ASCA, a large portion of the population could be at risk.

In the early 2000s, SCA attacks evaluation was suggested to be encompassed in cryptographic algorithm evaluation in many international
standards bodies, such as 3GPP security architecture \cite{3GPP}. However, due to a lack of testable methods
and practical tools, such an important suggestion never turned into practical standards and guidelines. 
There have been many academic attempts, but nothing led to standardisation. For instance, in a NIST report in 2011 \cite{gilbert2011testing}, a testing methodology was proposed to assess whether a cryptographic module utilising side channel analysis countermeasures can provide resistance to these attacks commensurate with the desired security
level. In a recent report \cite{abdulgadir2022side}, the authors developed and compared SCA-protected implementations of three finalists in the NIST LWC standardisation
process.
While there is no specific research dedicated to side channel attack standardisation, there have been industrial attempts to rectify some of the known attacks. For instance, in 2018 Google proposed a new technique to mitigate the infamous Spectre class of attacks. Similarly, Intel added hardware and firmware mitigations to tackle the same range of side channel attacks. Similarly, some general guidelines lines have been developed. For instance, the NSA TEMPEST includes acoustic emanations as a side channel but there are limitations in how they have defined acoustic in their terminology. Also, FIPS 140-3 draft, does not include acoustic emanations as a side channel, despite the fact that it has been used to extract RSA private keys from CPU's \cite{genkin2014rsa}.
Despite these efforts, there is no explicit standardisation work on ASC attacks. W3C specifications on sensors \footnote{w3.org/TR/generic-sensor/\#mitigation-strategies} (e.g., motion sensors on mobile devices) has a dedicated section to security and privacy considerations, where among the other risks, suggests keystroke monitoring as one of the possible threats enabled by such sensors. These sensors have proved to contribute to ASC attacks. 
The mitigation strategies suggest a range of methods, though none of them guarantees full support.  

In this paper, we present a practical fully--automated ASCA which deploys cutting edge deep learning models to improve the body of knowledge. We will address these research questions: (RQ1) Can we design and implement a fully automated ASCA pipeline, including the keystroke separation, feature extraction and predictions? (RQ2) Can we deploy an accurate deep learning approach for ASCA? (RQ3) Can we perform an accurate remote ASCA attack on VoIP communications considering the compression and information loss in the audio transmissions?

In this paper, we
contribute to the body of knowledge in a number of ways. (1) We propose a novel technique to deploy deep learning models featuring self-attention layers for an ASC attack on a keyboard for the first time.
    (2) We propose and implement a practical deep learning-based acoustic side channel attack on keyboards. We use self-attention transformer layers in this attack on keyboards for the first time.    (3) We evaluated our designed attack in real--world attack scenarios; laptop keyboards in the same room as the attacker microphone (via a mobile device) and laptop keystrokes via a Zoom call. We perform experiments and run multiple evaluations and our results outperform those of previous work.

\section{Related Work}
While they remain a relatively under-explored topic of research, ASCAs are not a new concept to the field of cybersecurity. Encryption devices have been subject to emanation-based attacks since the 1950s, with British spies utilising the acoustic emanations of Hagelin encryption devices (of very similar design to Enigma) within the Egyptian embassy \cite{5}. Additionally, the earliest paper on emanation-based SCAs found by this review was written for the United States' National Security Agency (NSA) in 1972 \cite{18}. This governmental origin of ASCAs creates speculation that such an attack may already be possible on modern devices, but remains classified. \cite{2} notes that classified documents produced by the NSA's side channel specification (TEMPEST) are known to discuss acoustic emanations. Additionally, the partially declassified NSA document NACSIM 5000 \cite{6} explicitly listed acoustic emanations as a source of compromise in 1982. 
Within the realm of public knowledge, ASCAs have seen varying success when applied to modern keyboards, employing a similarly varied array of methods. Surveying these methods, various observations may be made about the current research landscape.

In the last decade, the number of microphones within acoustic range of keyboards has increased and will likely continue to do so. In an attempt to explore these attack vectors, recent research has been utilising alternate methods of keystroke collection. As an example, in \cite{9}, the authors implemented an attack utilising a number of off-the-shelf smartphones. These devices (as is the case for a majority of modern phones) feature 2 distinct microphones at opposite ends of the phone. When used together, recordings made by the collective microphones provided sufficient time delay of arrival (TDoA) information to triangulate keystroke position, achieving over $72.2\%$ accuracy. \cite{11} built upon this research by implementing TDoA via a single smartphone in order to establish distance to a target device, eventually achieving $91.52\%$ keystroke accuracy when used within a larger attack pipeline. 

Alongside smartphones, video conferencing applications have seen promising results as an attack vector. Keystrokes intercepted from a VoIP call were used in \cite{10}, achieving a keystroke accuracy of $74.3\%$ and this success was echoed by \cite{16} which achieved a top-5 accuracy of $91.7\%$ via simply calling a victim over Skype. These successes mark the first ASCAs implemented without the need for physical access to a victim's vicinity and carry the implication that if a victim's microphone could be accessed covertly, a similar attack could be performed. 
The same implication can be found with the use of smartwatches as an attack vector. While it remains unlikely an attacker could covertly place their smartwatch in a private location such as an office, compromising a victim's smartwatch could allow unbridled collection of acoustic keystroke information. Additionally, smartwatches can uniquely access wrist motion, a concerning property which is utilised by \cite{12} to achieve $93.75\%$ word recovery.


One approach that saw prominent usage in the 2000's but has become less common in modern papers is the use of hidden Markov models (HMMs). A HMM (in this context) is a model trained on a corpus of text in order to predict the most likely word or character in the positions of a sequence. For example, if a classifier output `Hwllo', a HMM could be used to infer that `w' was in fact a falsely classified `e'. \cite{3} presents a method of ASCA attack on keyboards in which two HMMs are utilised: the first generating likely letters from a series of classes and the second correcting the grammar and spelling of the first. Similarly to \cite{3}, \cite{1} used a HMM to correct the output of a classifier and saw an increase from $72\%$ to $95\%$ accuracy when implemented. A difference in the two studies however, sheds light on a potential drawback to HMM usage (and the possible reason for lack of recent popularity). 


In much of the literature, neural networks are not perceived as very successful models when conducting keystroke recognition. In \cite{3}, a neural network was tested against a linear classifier and was deemed less accurate. Additionally, in \cite{7} a neural network was found to perform the worst out of all methods tested, and it is noted that neither \cite{3} nor \cite{7} could reproduce the results achieved in \cite{2} through use of a neural network.
\cite{10} found that multiple methods performed better than neural networks in testing, while \cite{30} implemented a neural network that performed third best out of all tested classifiers.
A majority of these papers give very little detail regarding the structure or size of the neural networks implemented, making comparison between them difficult, but in none of these cases was a neural network selected as the final model.
Given that Transformers were invented in 2018 by Vaswani et al. \cite{37}, this paper is the first use of neural networks featuring self-attention layers for an ASC attack on a keyboard.


Alongside models, variety exists between studies with respect to target devices. \cite{2}, the paper most commonly cited as the first ASCA targeting a keyboard, was written in 2004 and attacked high-profile plastic keyboards synonymous with the time. Despite being such an early paper in the field, success was found in attacking an ATM keypad, a corded telephone as well as 2 keys from a laptop keyboard.
While \cite{3} and \cite{4} perform their experiments on keyboards similar to those from \cite{2}, \cite{7} investigates a more modern keyboard with a slightly recessed design. The keycaps remain large and plastic however and differ greatly from modern laptop keyboards. \cite{7}'s authors do however acknowledge that the testing of laptop keyboards may produce different results, due to a lack of `release peak' in the waveform.


Of the surveyed literature, \cite{11} and \cite{16} were the only 2 papers to feature ASCAs on full laptop keyboards and are the most promising studies with respect to real-world implementation. Both papers utilise two statistical models used in similar ways: the first to infer some information regarding the victim's environment and the second to classify keystrokes into letters. The two papers differ in most other ways however, with \cite{16} gathering keystrokes via Skype and the inbuilt microphone of the laptops, while \cite{11} utilises a mobile phone placed near the victim's computer. Additionally, \cite{16} uses k-NN clustering and a Logistic Regression classifier while \cite{11} utilises support vector machines (SVMs). Despite their differences, both papers are notable for their accuracy, with \cite{11} achieving $91.2\%$ in cross validation and $72.25\%$ when attacking unknown victims and keyboards. Meanwhile \cite{16} achieves a top-5 accuracy of $91.7\%$ given knowledge of the victim's typing style. \cite{11} implements it's attack on 2 laptops, made by Alienware and Lenovo respectively and is notable for being the only study to feature membrane keyboards. \cite{16} presents a much more representative study of keyboards, attacking 6 laptops, two of each: MacBook Pro 13" 2014, Lenovo Thinkpad E540 and Toshiba Tecras M2.

\section{Attack Design}

In this section, we discuss the overall design of our proposed ASC attack. Then, we explain our proposed approach in data collection, feature extraction and our model design.

\subsection{Fully--automated On--site and Remote ASCA}

In both set of experiments (via phone and Zoom), 36 of the laptop's keys were used (0-9, a-z) with each being pressed 25 times in a row, varying in pressure and finger, and a single file containing all 25 presses. 

\begin{figure}
    \centering
    \includegraphics[width=8.7cm]{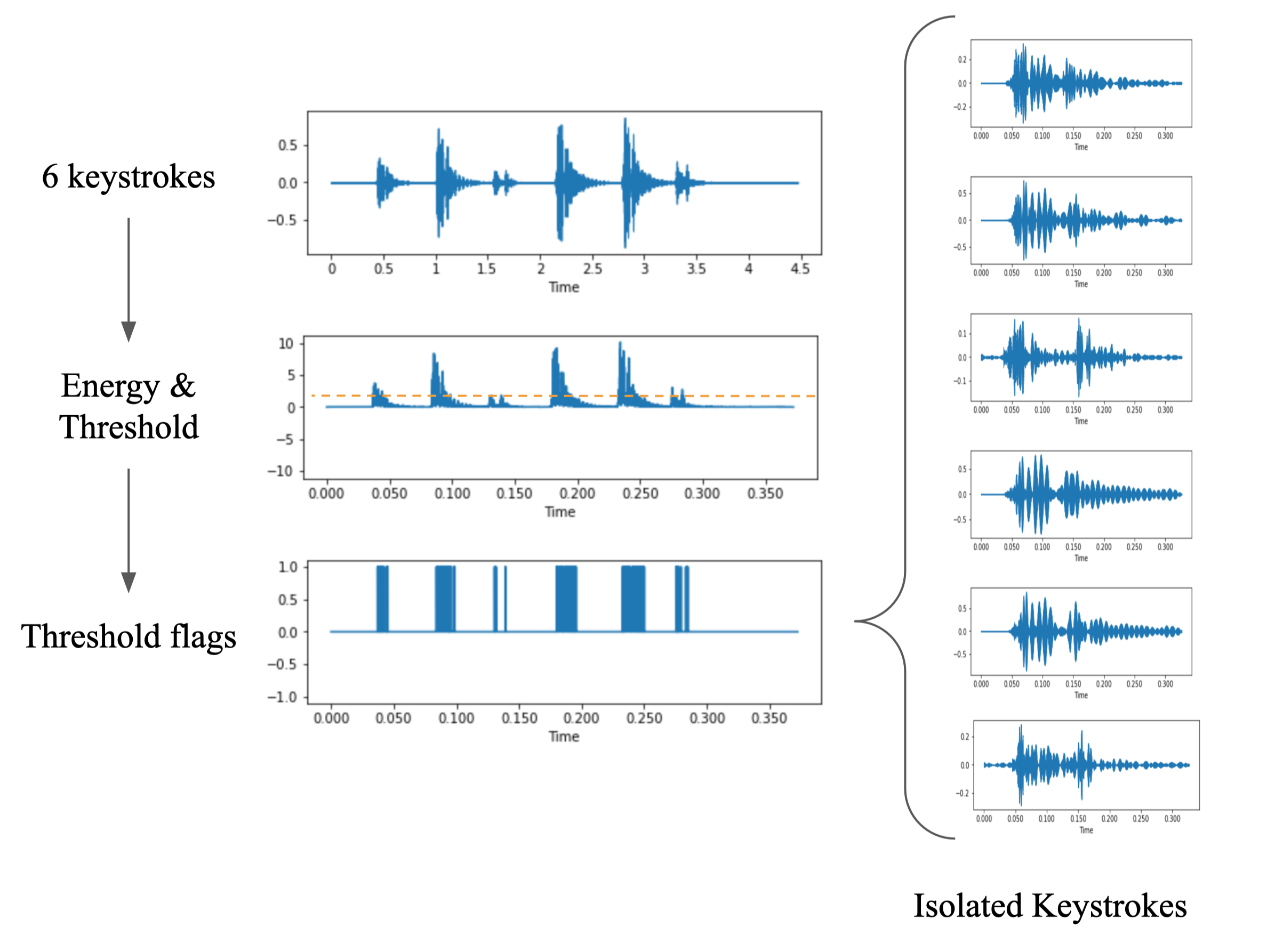}
    \caption{Keystroke isolation process, signals are converted to energy via FFT, then flagged when crossing the threshold to mark keystrokes}
    \label{fig:energy}
\end{figure}

\textbf{Keystroke isolation:} Once all presses were recorded, a function was implemented with which individual keystrokes could be extracted. Keystroke extraction is executed in a majority of recent literature \cite{3,4,7,11} via a similar method: performing the fast Fourier transform on the recording and summing the coefficients across frequencies to get `energy'. An energy threshold is then defined and used to signify the presence of a keystroke. The complete isolation process can be seen executed on an excerpt from the phone data in Fig. \ref{fig:energy}. The keystrokes isolated for this data were of fixed length 14400 (0.33s).

Isolating the keystrokes proved more difficult with the Zoom data set. Given the noise suppression present in the Zoom recording, the volume of keystrokes varied massively, making the setting of a threshold value difficult. To bypass this, a loop was implemented in which the threshold was adjusted by increasingly small values until the correct number of keystrokes was found, shown in algorithm \ref{alg:loop}.

\textbf{Feature extraction:} Multiple methods of audio feature extraction exist and the literature commonly varies as to which is used, however there are common candidates across most ASCA studies including (i) the Fast Fourier Transform (FFT)\cite{2},
(ii) Mel-Frequency Cepstral Coefficients (MFCC)\cite{1,3,10,11,30}, and 
(iii) Cross Correlation (XC)\cite{4,15}.
In this paper, we propose the use of mel-spectrograms as a method of feature extraction for a DL model. A mel-spectrogram is a method of depicting sound waves and is a modified version of a spectrogram. Spectrograms represent sound as a map of coloured pixels, with the Y axis representing frequencies and the X axis representing time. The brightness of a pixel $(x,y)$ in a spectrogram represents the amplitude of a frequency ($y$) at a given time ($x$). This concept is then built-upon to form mel-spectrograms, in which the unit of frequency is adjusted to mels: a logarithmic scale more representative of how humans hear sound. 
Through some initial experiments, we found out that it spectrograms (and specifically mel-spectrograms) represent sound in a visually recognisable manner, a valuable property when considering that DL has been found repeatedly to be one of the best approaches for image classification. For example, in 2021 CoAtNet achieved a state-of-the-art accuracy of $90.88\%$ when classifying 1,000 ImageNet images \cite{29}.
It is worth consideration that both FFT and MFCC produce similarly visual depictions of features. In the case of FFT, a spectrogram with linear axis in both frequency and volume is produced. This property makes FFT less suitable for this paper, since a majority of features in keystroke sounds are within the lower frequencies \cite{15,10,2} and would therefore be less distinguishable on a linear scale. 
Meanwhile, MFCC involves performing the discrete cosine transform on a mel-spectrogram, producing a compressed representation that prioritises the frequencies used in human speech. Since, for this paper, human speech is not the target, and the removal of frequencies could risk the loss of relevant data, MFCC was decided to be less suitable than mel-spectrograms.

\textbf{Data augmentation:} Prior to feature extraction, signals were time-shifted randomly by up to $40\%$ in either direction. This time shifting is an instance of data augmentation, in which the amount of data input to a DL model is artificially increased by slightly adjusting existing inputs \cite{39}. 
The mel-spectrograms were then generated using 64 mel bands, a window length of 1024 samples and hop length of 500 (255 for the MacBook keystrokes, given their shorter length), resulting in 64x64 images.
Using the spectrograms, a second method of data augmentation was implemented called  masking. This method involves taking a random $10\%$ of both the time and frequency axis and setting all values within those ranges to the mean of the spectrogram, essentially `blocking out' a portion of the image. 
Using time warping and spectrogram masking combined is called SpecAugment and was found to encourage the model to generalise and avoid overfitting the training data \cite{40,38}.


\begin{figure}[t]
    \centering
    \includegraphics[scale = .15]{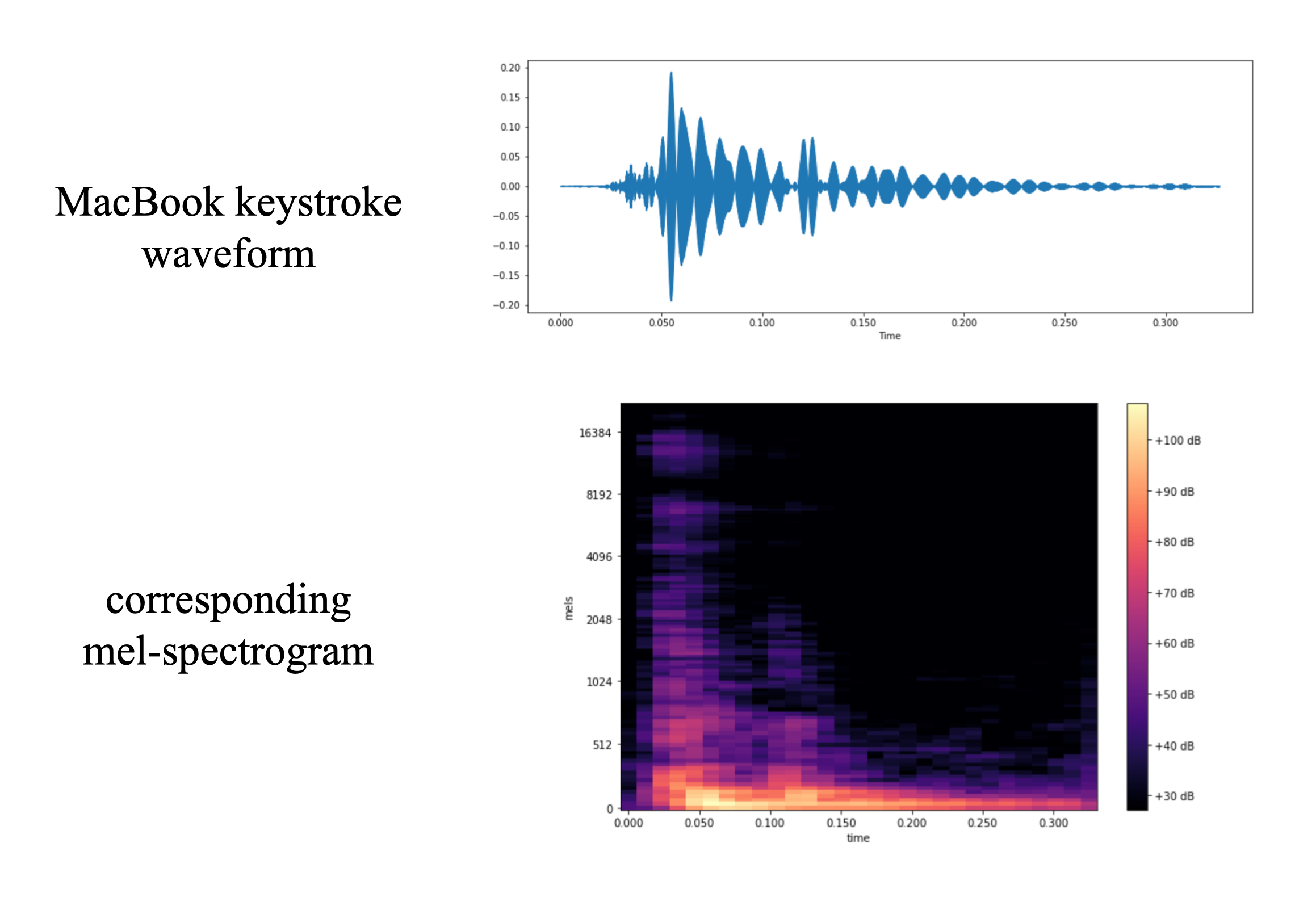}
      \includegraphics[scale = .2]{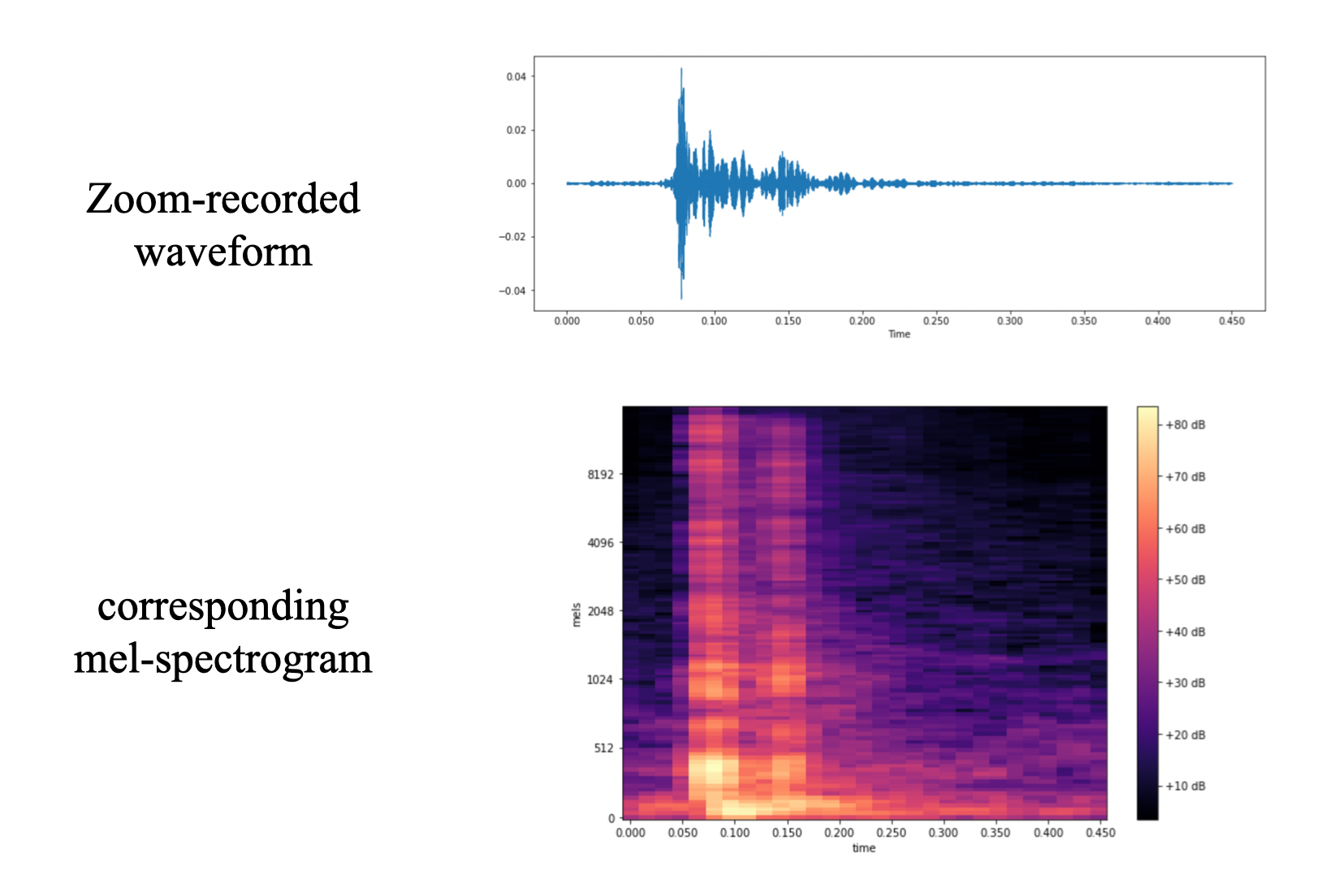}
    \caption{Waveform and corresponding mel-spectrogram of Left: Phone recording, and Right: Zoom recording.}
    \label{fig:MBPmelwav}
\end{figure}


\begin{algorithm}
\label{alg:loop}
\SetAlgoLined
\KwResult{A set of isolated keystrokes, $S$}
 \textbf{Input:} A function for isolating keystrokes $Iso(F, P)$, an initial prominence threshold, $P$, a recording of keystrokes, $F$, a step value, $s$ and a target number of keystrokes, $T$\;
 \textbf{Initialisation:} An empty list of keystrokes, $S = \{\}$\;
 \While{$S.length\neq T$}{
    $S = Iso(F, P)$
    
    \If{$S.length < 25$}{
    $P = P - s$
    }
    \If{$S.length > 25$}{
    $P = P + s$
    }
    $s = s*0.99$
 }
 \caption{Zoom keystroke threshold setting}
\end{algorithm}

Having converted keystrokes from each data set into a more visual medium, more direct comparisons could be made.
MacBook keystrokes (similar to the keystrokes examined in the literature \cite{2,3,11}) have only 2 visible peaks: the `push' and `release' peaks respectively. 
The 2 peak structures shown in Fig. \ref{fig:MBPmelwav} are similar to each other, implying that such a structure is native to the MacBook keyboard regardless of recording method, a noticeable difference however is the large range of frequencies present in the zoom recording. 
The Zoom peaks extend much higher than that of the phone-based recordings, indicating significant data in multiple frequencies that were not present when recorded via phone. 

\begin{figure}
    \centering
    \includegraphics[scale=.65, angle =0 ]{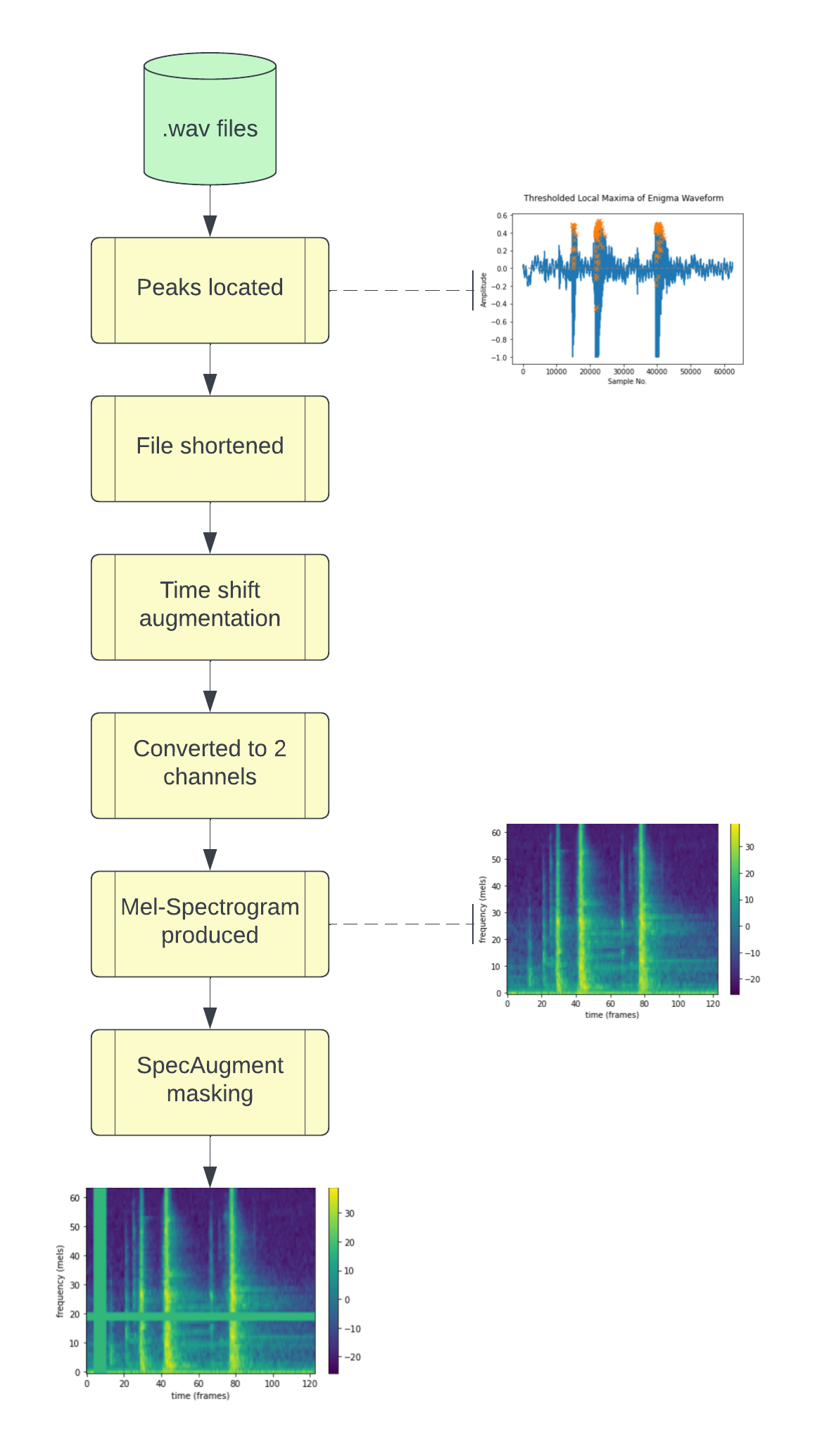}
    \caption{The data processing pipeline.}
    \label{fig:flow}
\end{figure}

The overall data preparation procedure for our data was inspired by the structure presented in \cite{38} and is shown in Fig. \ref{fig:flow}.

\subsection{Model Selection and Implementation}
We implemented a deep learning model on the processed data. Given the visually-distinguishable nature of mel-spectrograms, a model proven to work well for image classification was needed. We chose the CoAtNet model created in \cite{29} due to its excellent performance on the ImageNet classification data set and it's far lower training time compared to similarly performing models. 
Generally describing, CoAtNet can be seen to consist of two depth-wise convolutional layers followed by two global relative attention layers. The act of combining convolution and self-attention methods allows for rapid processing of patterns in the data while down-sampling the size (convolution) before determining the relevance of these patterns to one another through the calculation of attention scores (self-attention) \cite{29}.
We used PyTorch \cite{34} to implement an instance of CoAtNet. 

The code presented in \cite{41} returns probabilities given in dimensions far greater than the desired shape and number of classes required for this work. To overcome this implementation issue, the output of CoAtNet was reduced to a percentage probability relating to each of the keys. The output of the final self-attention layer was therefore subjected to a 2D average pool followed by a fully-connected linear layer. These additions not only produced the desired output, but also more closely reflect the desired implementation structure of CoAtNet presented in \cite{29}.
The initial parameters used for the implemented models were based on those from the original study including the use of the Adam optimiser and cross entropy loss criterion \cite{29}. With the inputs being of identical size, the only parameter requiring adjustment between data sets was the output of the final fully connected layer.




When implementing a DL model such as CoAtNet, it is important to establish values for various hyperparameters that will define specific behaviours of the model. While some hyperparameters have values found commonly within the literature, a majority of hyperparameter combinations simply have to be tried and validated in order to be compared. One common approach to the hyperparameter optimisation problem is grid search, in which all combinations are tested and the best selected. This method, and those like it, take a large amount of time and, given the already complex model implemented in this paper, repeatedly training and evaluating models would place limitations on the validity of a real-world attack. Ideally, a minimal amount of adjustment would be required to hyperparameters in order to produce a satisfactory model, requiring less time and less competent hardware to execute.

In order to achieve satisfactory model performance with minimal overhead, three hyperparameters were experimented with for each of the three models. The selected hyperparameters were: maximum learning rate (LR), total training epochs and the method of splitting data. The first, LR, represents the rate at which a model's weights adjust to suit the training data, the second defines how many times the model is trained on the entire training data set and the third represents the method used to divide the test, training and validation sets from the overall data pool.

In order to find appropriate values of these hyperparameters, models were trained then tested against the validation data. 
In order to train the DL model, we first normalised the data and input it to the model in batches. The model then produced class probabilities for each item in the batch. 
    These probabilities were used to calculate cross entropy loss and accuracy, with respect to the true values.
    Then the loss was used by the optimiser to perform the backpropagation algorithm and adjust the model to better suit the true values. Next, 
    The scheduler was stepped to reduce the learning rate. And finally, 
    every 5th epoch, the model was switched to an evaluation configuration and tested on the validation data.
Utilising such a procedure, models were assigned an LR of 1e-3 and trained for varying numbers of epochs on a stratified split of the data. For each total number of training epochs, the highest validation accuracy seen is logged in table \ref{table:Epochs}.

\begin{table}
\caption{The best validation set accuracy. LR was set to 1e-3 and all models were trained on a stratified split of data.}
\label{table:Epochs}
\centering
\footnotesize
\begin{tabular}{l|l|lllll}
\hline
Total Epochs & &1300 &1100 & 500 & 300 & 100 \\\hline
Peak Validation &Phone & 0.87 & 0.89 & \textbf{0.92} & 0.46& 0.29\\
Accuracy &Zoom& 0.26 & \textbf{0.52}&0.14&0.32&0.18\\\hline
\end{tabular}
\end{table}



\begin{table}[t]
\renewcommand{\arraystretch}{1.3}
\caption{Peak validation accuracy achieved when using varying values for all hyperparameters. LR = Learning Rate, PVA = Peak Validation Accuracy}
\label{table:Splits2}
\centering
\begin{tabular}{p{1cm}p{1cm}p{1cm}p{1cm}p{1.4cm}}
\hline
\textbf{Data} & \textbf{Split} & \textbf{LR} & \textbf{Epochs} & \textbf{PVA} \\
\hline
Phone & Random & 5e-4 & 1100 & \textbf{0.96 }\\
 & Random & 1e-3 & 500 & 0.92 \\
 & Stratified & 1e-3 & 500 & 0.92 \\
 & Stratified & 1e-3 & 1100 & 0.89 \\
 \hline
Zoom & Random & 5e-4 & 1100 & \textbf{0.96 }\\
 & Random & 1e-3 & 1100 & 0.59 \\
 & Stratified & 1e-3 & 1100 & 0.52 \\
 & Stratified & 5e-4 & 1100 & 0.44 \\
\hline
\end{tabular}
\end{table}

While the number of epochs required to train a model to a point of convergence varies throughout the literature, the results of this experiment show that this classification requires an uncommonly high number of training epochs when using the default values for LR, momentum (a hyperparameter for the Adam optimiser) and other hyperparameters. The implication of this is that given these default values, the model acquires meaningful interpretation of the data at a relatively slow rate. 
A second preliminary experiment was undertaken in which the method of splitting the data was varied. The purpose of this experiment was to provide insights as to the impact of uneven data sets on model performance, as an inconsistent data set is more analogous to a real-world attack. For each dataset, a model was trained for 1100 epochs on an identically-sized training set, split by a either a seeded random or stratified method. The models were then tested for accuracy on similarly-split validation data. 
We observed that differing the splits of data made little difference to the peak validation set accuracy. Of the three types of data, the Zoom dataset saw the largest difference, increasing validation accuracy by over $13\%$ when using the random split. The other two types of data saw little difference between when trained on different splits, a result that is not surprising. Given the even distribution of classes in all three data sets, random samples of these data sets are expected to be similarly distributed, despite some variance, leading to mostly similar data sets regardless of split. 
It remains as potential further research as to how models would perform on vastly different distributions of training or testing data.

\begin{figure*}
    \centering
    \includegraphics[scale=.4]{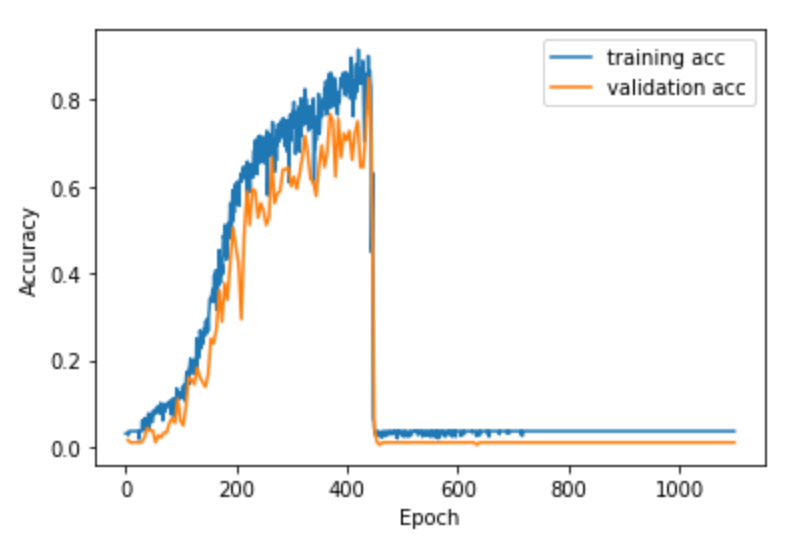}
    \includegraphics[scale=.4]{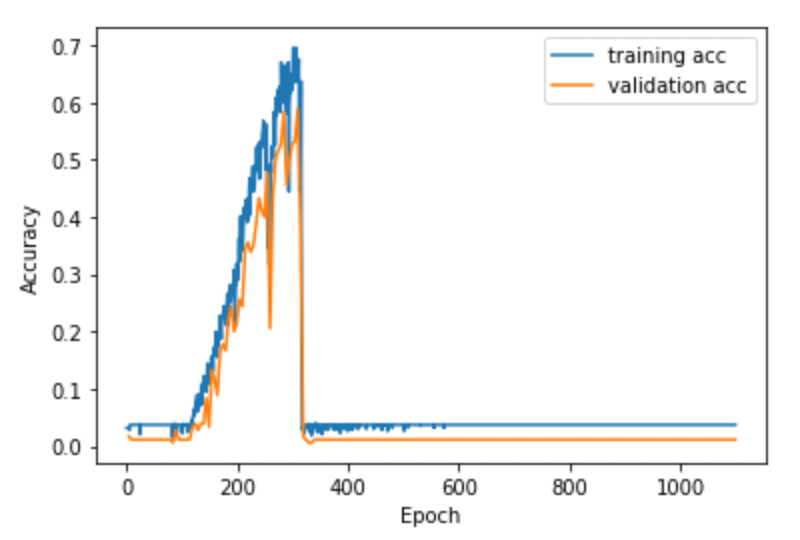}
    \caption{The training and validation accuracy, 1100 epochs, LR = 1e-3 on randomly split data, left: Phone, Right, Zoom.}
    \label{fig:PhoneErr}
\end{figure*}

Following the execution of this experiment, the training and validation accuracy was plotted for each of the models to inspect for anomalies, convergence and other behaviours relevant to DL model performance. When examining the accuracy of both MacBook models when trained on randomly split data, an abnormality was found. 
This abnormality can be seen in Fig. \ref{fig:PhoneErr}. 

\begin{figure*}
    \centering
    \includegraphics[scale=.4]{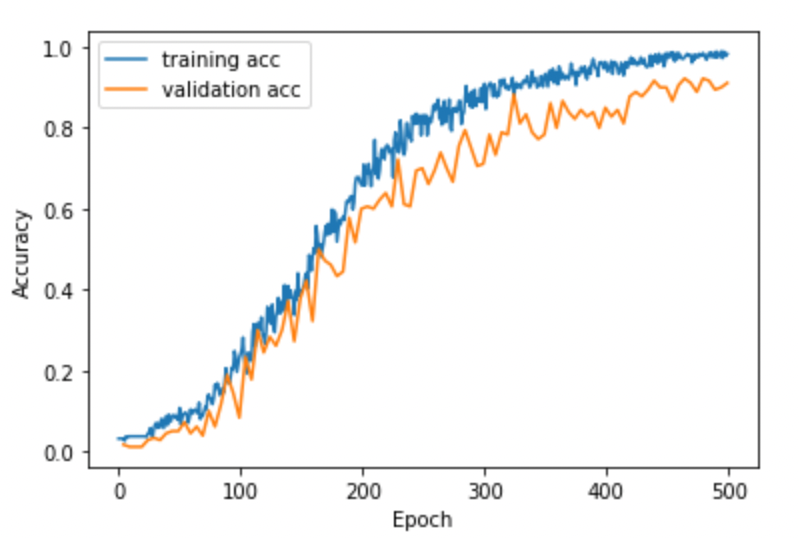}
        \includegraphics[scale=.4]{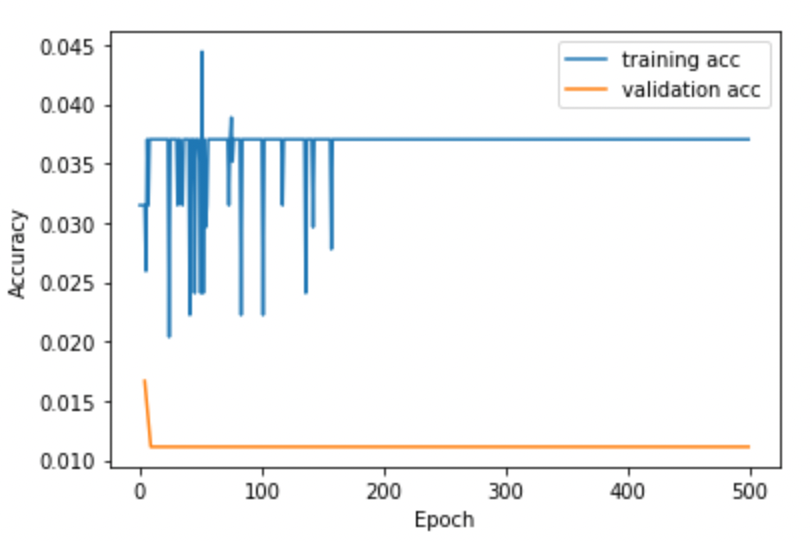}
    \caption{The training and validation accuracy, 500 epochs, LR = 1e-3 on randomly split data, Left: Phone, Right: Zoom.}
    \label{fig:500PhoneErr}
\end{figure*}

As can be seen, the models displayed good training progress for a period of 300-400 epochs, before a sudden `reset' to entirely random prediction. This pattern, while initially achieving a good performance on validation data, clearly indicated a flaw in one or more parameters of the classifiers. 
To overcome this issue, both models were trained again on the random split, but for 500 epochs as opposed to 1100, in an attempt to avoid training them past this point of `collapse'. The results are shown in Fig. \ref{fig:500PhoneErr}. Here, while the phone data classifier showed some convergence, as well as a slightly improved peak validation accuracy, the Zoom classifier showed no sign of convergence and performed consistently worse than random on the validation data across all training epochs. 


The next step in attempting to address this implementation issue was to adjust the learning rate as opposed to the number of training epochs. In this experiment, each model was trained for the full 1100 epochs, but with an initial learning rate of 5e-4, half the default value. 


We concluded that adjusting the learning rate was sufficient to address this issue. By slowing the rate of learning but allowing the models to train for the same number of epochs, not only do they avoid `resetting' to random, but they also train to a greater accuracy than achieved previously. Following this insight, all 3 parameters were tested in various combinations, with the results reported in Table \ref{table:Splits2}.

From this table, it can be seen once again that for the 
Phone classifier, the split of data does not appear to be relevant. Meanwhile, the Zoom classifier saw far greater performance when trained on a random split of data with a lower learning rate. Additional values of epoch and learning rate were experimented with, however a validation accuracy of $0.52$ could not be improved upon for the stratified Zoom data.
The reason for the Zoom classifier's performance difference across the two splits could be explained by potential anomalous features in the data. Should the recording of a certain key have been effected disproportionately by Zoom's noise reduction feature, inclusion of more instances of that key in the input data could cause confusion in the model's training.
The process of overcoming this issue and consequently experimenting with the three chosen hyperparameters allowed for selection of hyperparameter values for the final models, as presented in Table \ref{table:paramsMBP}.


\begin{table}[h]
\renewcommand{\arraystretch}{1.3}
\caption{Default hyperparameters used for the model and data processing when training the MacBook keystroke classifiers}
\label{table:paramsMBP}
\centering
\small
\begin{tabular}{p{4.15cm}p{3.15cm}}
\hline\
\textbf{Parameter} & \textbf{Value} \\
\hline
Epochs & 1100 \\
Batch Size & 16 \\
Loss Type & Cross Entropy \\
Optimiser & Adam \\
Max Learning Rate & 5e-4 \\
Annealing Schedule & Linear \\
Timeshift Percentage & 0.4 \\
Max Mask Percentage & 0.1 \\
Number of Masks Per Axis & 2 \\
Mel Bands & 64 \\
FFT Window Size & 1024 \\
Hop Length & 225 \\
Data Split & Random \\
Normalised Data & Yes \\
\hline
\end{tabular}
\end{table}


\begin{figure*}
    \centering
    \includegraphics[scale=.25]{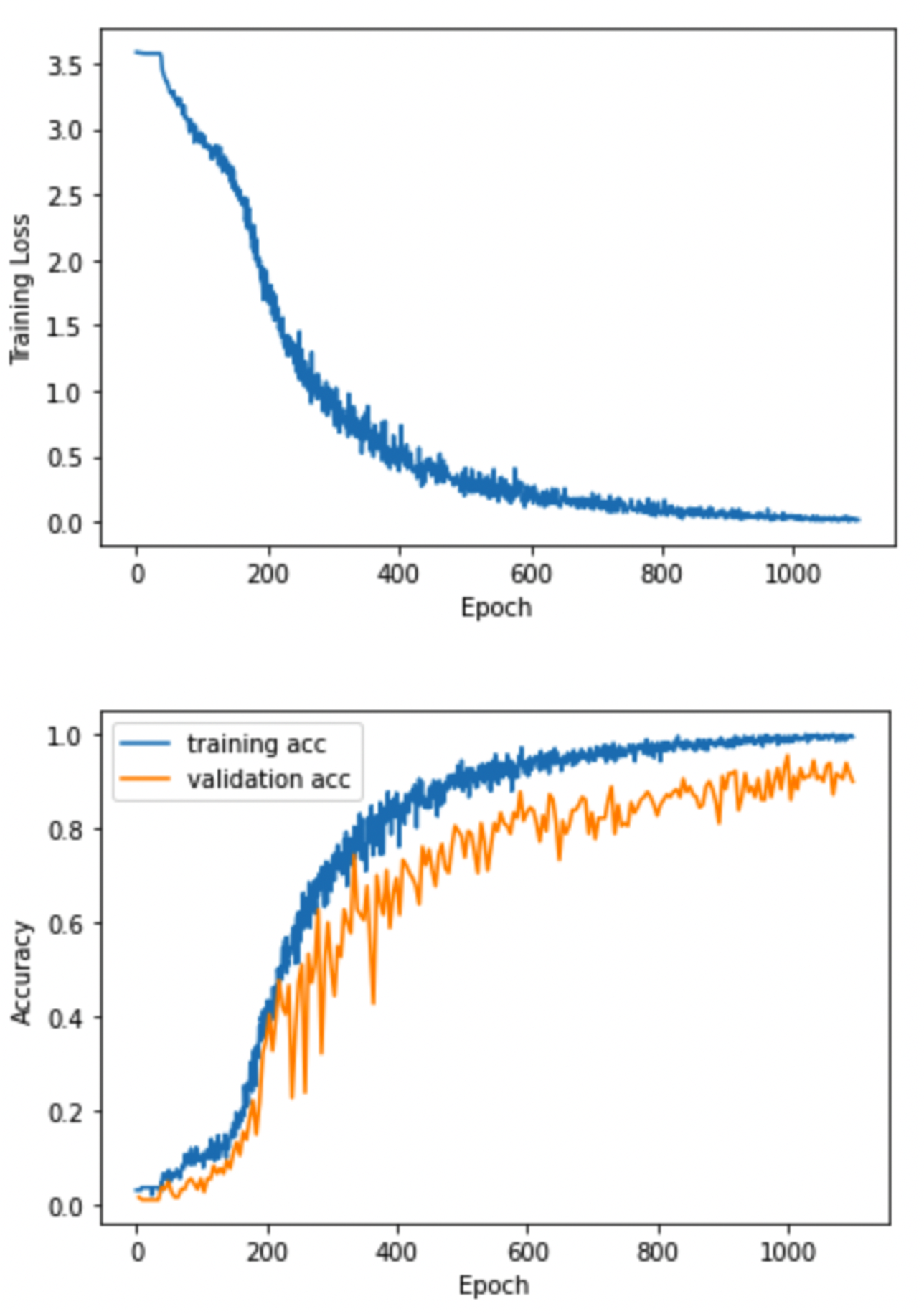}
    \includegraphics[scale=.25]{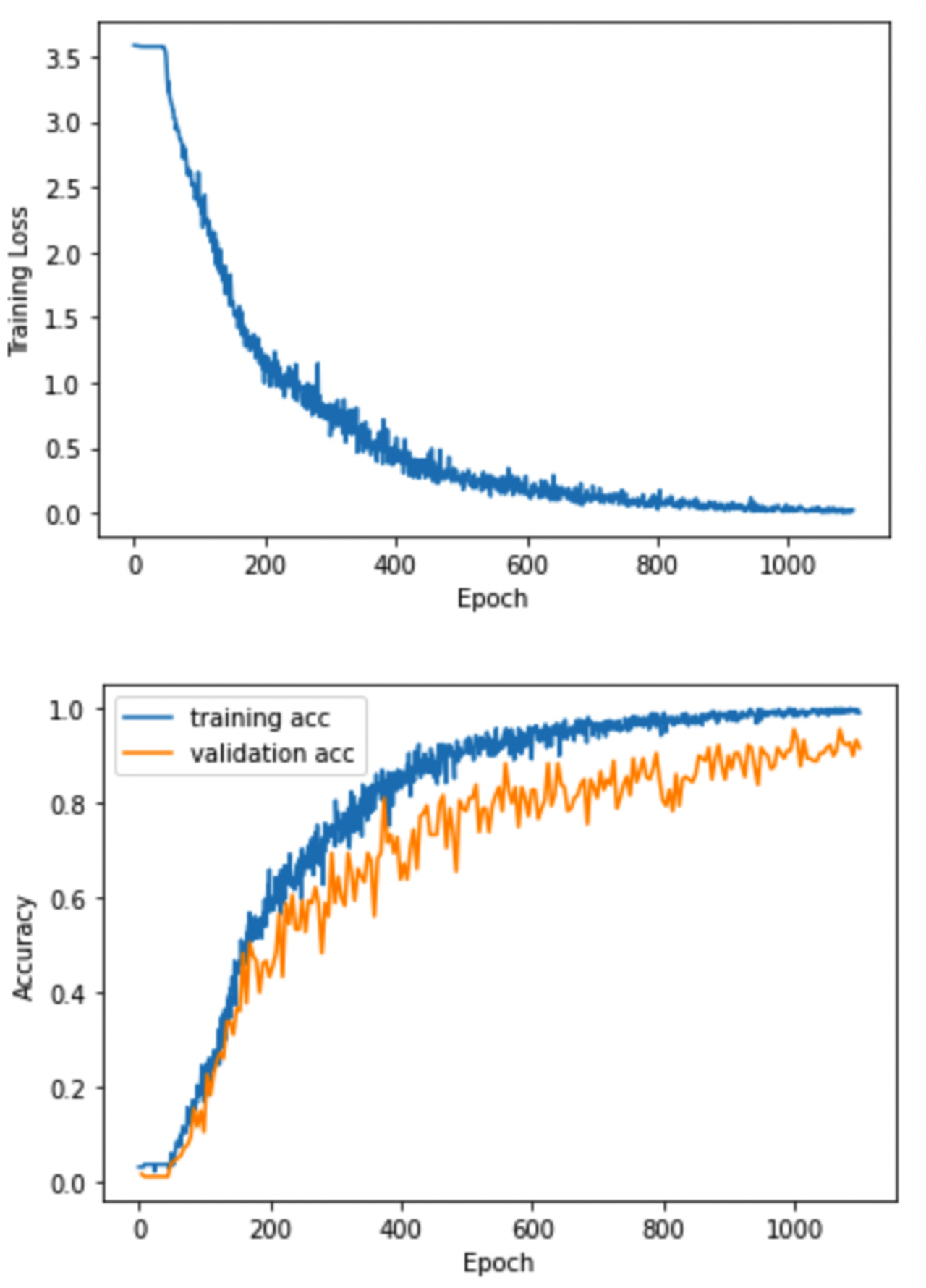}
    \caption{The testing accuracy, validation accuracy and training loss plotted relative to epochs, Left: Phone, Right: Zoom}
    \label{fig:MBPEval}
\end{figure*}

Having determined the hyperparameter values to be used, models were instantiated with the desired values and trained on their respective data sets. As in the preliminary experiments, every 5 epochs models were tested on the validation data and the resulting accuracy values were plotted in Fig.\ref{fig:MBPEval}. 
Throughout training, cross entropy loss was calculated from the output class probabilities, this loss can be seen to reduce across training epochs in the same figures.

The methodology presented throughout this section is validated by both concurrent and face validity. The evaluation approaches described in this subsection have considerable face validity given the metrics selected (f1-score, precision, recall) being objective measurements of a model's performance on a given test set. 
Concurrent validity refers to validity inherited through relation to existing validated tests. In such a manner, the methodology used throughout this paper inherits a considerable amount of validity from the existing literature. While specifically mel-spectrograms were used as input features for this paper, the process of creating mel-spectrograms uses, and is used in, the creation of FFT and MFCC features respectively. Both of these methods are employed successfully throughout the literature \cite{2,1,3,10,11,30,4,15}, lending validity to the use of mel-spectrograms as features. Additionally, the model architecture CoAtNet has seen successful use in image classification \cite{29} and in it's structure uses self-attention layers, an increasingly active component in DL research. 
As a metric, accuracy is prevalent across the literature \cite{1,2,3,4,7,9,10,11}, as is top-x accuracy (top-5 in this case) \cite{2,4,10,11}. 

%

\section{Experiments and Results}
In this section, we presents our preliminary experiments for our classifier configurations and the results. We also present the final results of our evaluation for these acoustic attacks. 

\subsection{Sample Collection}


We ran our experiments on a MacBook Pro 16-inch (2021) with 16GB of memory and the Apple M1 Pro processor, which is a popular off-the-shelf laptop. 
This laptop features a keyboard identical in switch design to their models from the last 2 years and potentially those in the future. Additionally, the small number of available models at any one time (presently 3, all using the same keyboard) means that a successful attack on a single laptop could prove viable on a large number of devices. 

\begin{figure}
    \centering
    \includegraphics[scale=.04]{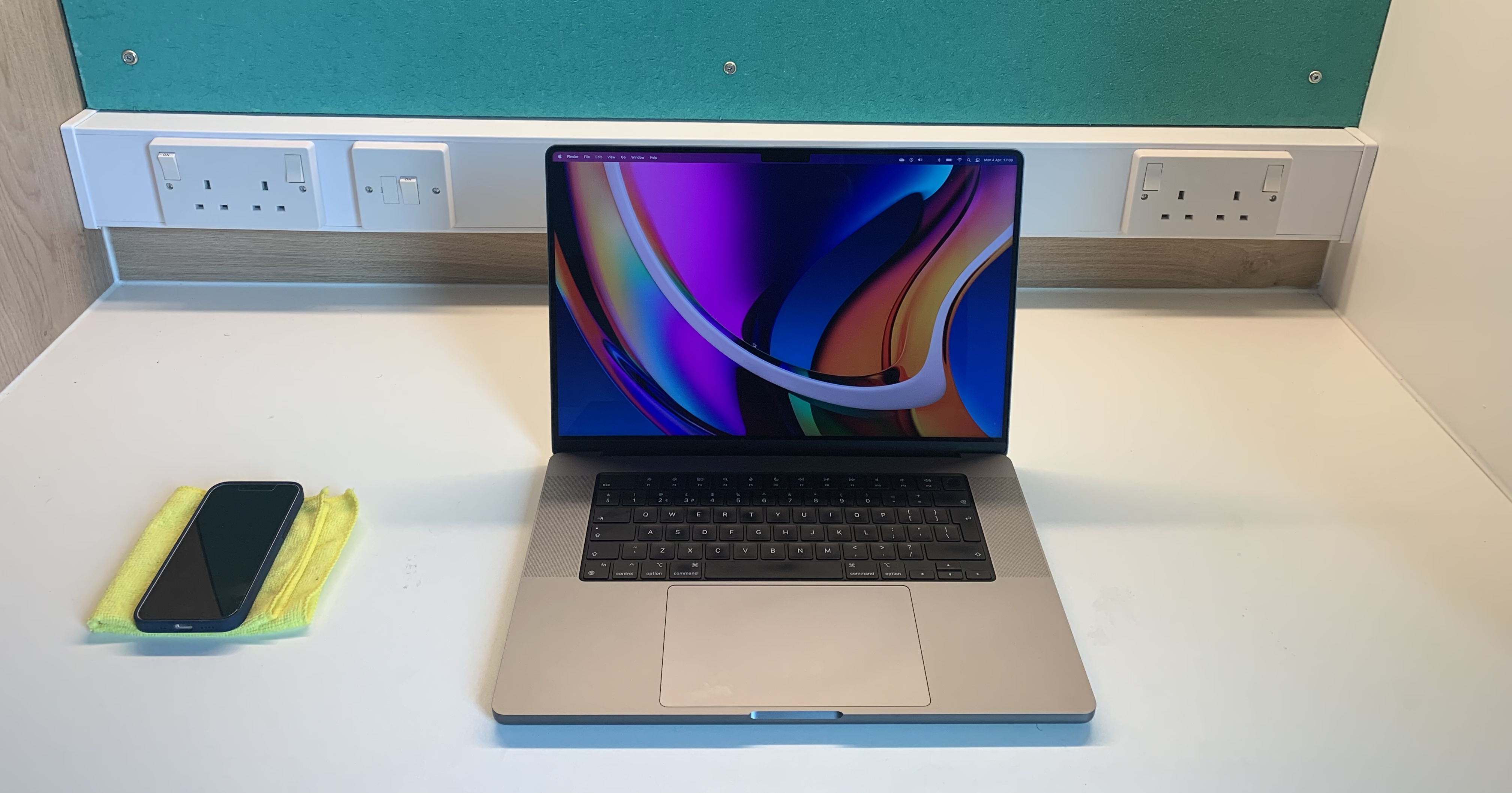}
    \caption{Desk setup for recording keystrokes}
    \label{fig:desk}
\end{figure}

\textbf{Phone-recording mode:} We used and iPhone 13 mini placed 17cm away from the leftmost side of the laptop on a folded piece of micro-fibre cloth (shown in Fig. \ref{fig:desk}). The purpose of the cloth was to remove some desk vibration in the recording (as this would vary based on the type of desk used), instead encouraging the model to learn primarily from acoustics. Recordings were made in stereo with a sample rate of 44100Hz and 32 bits per sample. 

\textbf{Zoom-recording mode: } For the second laptop dataset (referred to as `Zoom-recorded data'), keystrokes were recorded using the built-in function of the video conferencing application Zoom. The Zoom meeting had a single participant (the victim) who was using the MacBook's built-in microphone array. The noise-suppression parameter of Zoom was set to the minimum possible (`low') but could not be completely turned off. Before typing, the `Record on this Computer' button was pressed and after logging keystrokes the `stop' button was pressed, producing a .m4a sound recording that was converted to .wav format. As noted in \cite{10} and \cite{16}, recording in this manner required no access to the victim's environment and in this case, did not require any infiltration of their device or connection.

\subsection{Final Results}
In order to report our final results, confusion matrices and classification reports were created for each of the classifiers based on test set performance.
A confusion matrix depicts the number of times a model classified an instance of each class on the $X$-axis as being in a class on the $Y$-axis. For example, a value of 3 in cell $(x,y)$ where $x=1$ and $y=2$ would mean the classifier output 2 for an instance of class 1, 3 times. Therefore, a perfect classifier would result in a confusion matrix with 0 in every cell where $x\neq y$.
A classification report details the precision, recall, f1-score and support for each class in the data set. Given the large number of classes in all three datasets, we present just the averaged values of these metrics across all classes, as well as the support of the entire test set.



\begin{figure}
    \centering
    \includegraphics[width=8.7cm]{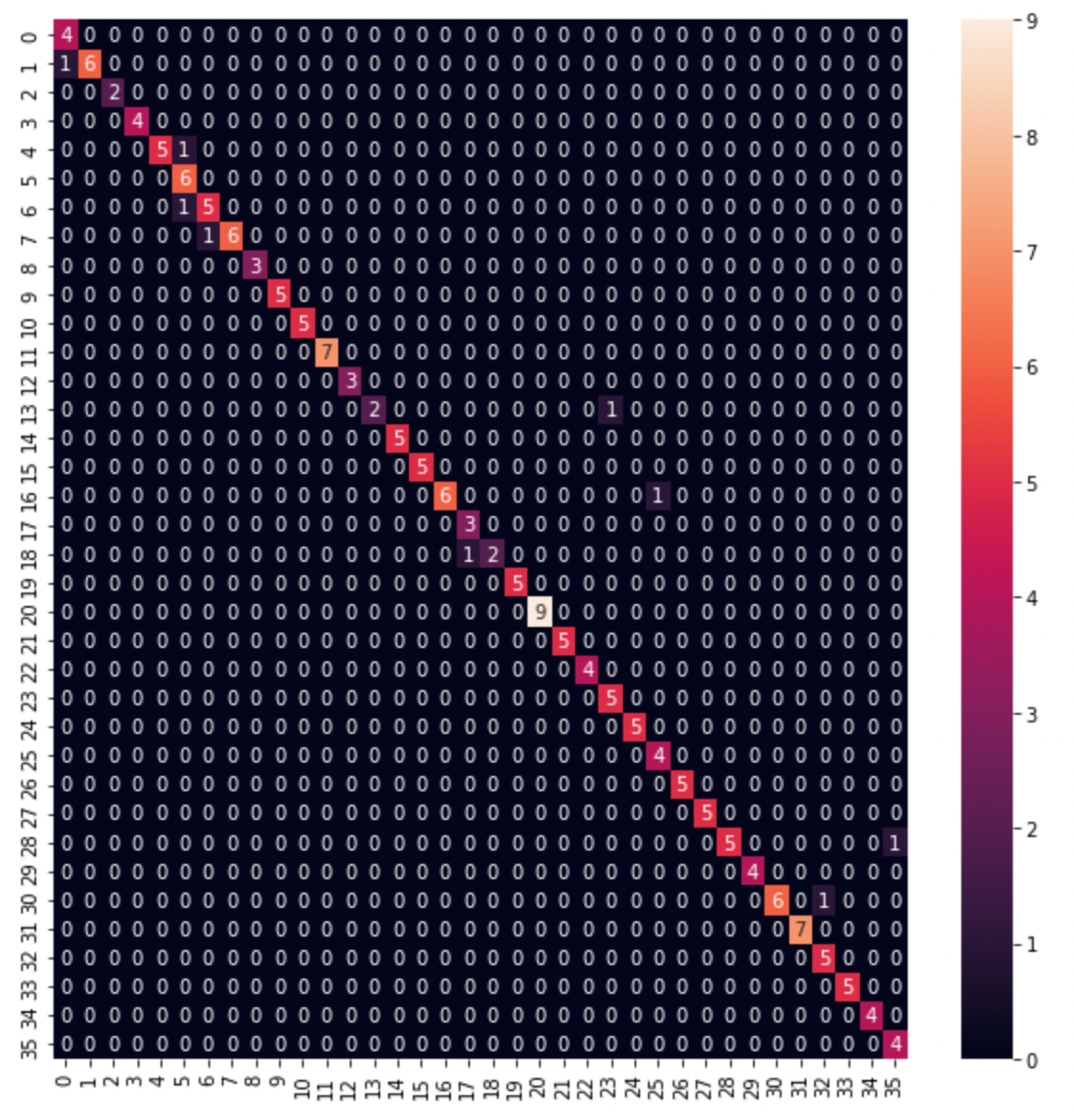}
   \caption{Confusion matrix for the phone-recorded MacBook keystroke classifier when evaluated on unseen test data}
    \label{fig:MBPCM}
\end{figure}

\begin{table}[t]
\renewcommand{\arraystretch}{1.3}
\caption{Classification report for the phone-recorded MacBook keystroke classifier}
\label{table:MBPCR}
\centering
\begin{tabular}{l|llll}
\hline
 & \textbf{Precision} & \textbf{Recall} & \textbf{F1-Score} & \textbf{Support} \\
\hline
Accuracy & \_ & \_ & 0.95 & 180 \\
Macro Avg & 0.96 & 0.95 & 0.95 & 180 \\
Weighted Avg & 0.96 & 0.95 & 0.95 & 180 \\
\hline
\end{tabular}
\end{table}

\begin{table}[t]
\renewcommand{\arraystretch}{1.3}
\caption{Classification report for the Zoom-recorded MacBook keystroke classifier}
\label{table:ZoomCR}
\centering
\begin{tabular}{l|llll}
\hline
 & \textbf{Precision} & \textbf{Recall} & \textbf{F1-Score} & \textbf{Support} \\
\hline
Accuracy & \_ & \_ & 0.93 & 180 \\
Macro Avg & 0.94 & 0.94 & 0.94 & 180 \\
Weighted Avg & 0.94 & 0.93 & 0.93 & 180 \\
\hline
\end{tabular}
\end{table}

This notion of position playing a large part in keystroke recognition is reinforced by the tendency for false-classifications to be only a single key `away' in the phone recording classifier's confusion matrix. In Fig. \ref{fig:MBPCM}, $5/9$ false-classifications are a single key away from the true value, and $6/9$ are within 2 keys of the true value, and hence show consistent patterns. 
while the MacBook's keystrokes were found to be slightly easier to classify, both keyboards rely on a similar positional layout and are therefore similarly classifiable when recorded with external microphones. With regards to overall susceptibility to an attack, the quieter MacBook keystrokes may prove harder to record or isolate, however, once this obstacle is overcome, the keystrokes appear to be similarly if not more susceptible.


While the data used for both MacBook classifiers was created for this paper, and consequently no other studies have been undertaken on it, the models implemented in this paper were found to have a higher accuracy than all other neural networks surveyed in the literature. Additionally, \cite{16} achieved a top-5 accuracy of $91.7\%$ when classifying MacBook keystrokes via the video conferencing application Skype. When trained on keystrokes from a similar laptop, recorded via a similar medium, the Zoom keystroke classifier presented by this paper achieved an accuracy of $93\%$.


Of the two recording modes, the precision recall and f1-score were higher by 0.02 for the model trained on the phone data. Such a small difference implies that the utilisation of alternative methods of recording may not necessarily diminish classification accuracy by a noticeable amount. It remains as a potential direction of future research as to whether other discrete methods of recording maintain a similar effectiveness.
Both the phone and Zoom recording classifiers achieved state-of-the-art accuracy given minimal training data in a random distribution of classes. In addition, this paper has shown the effectiveness of both mel-spectrograms as features and self-attention transformers as models when performing keystroke classification. The implementation of such models and feature extraction methods could be used alongside or to replace existing methods in order to further the field.

An observation from the results supporting the possibility of a real-world ASCA is the tendency of each classifier to cluster false-classifications around the correct key. This trait was recognised in \cite{30} and implies that a false-classification may still hold information regarding the location of the true key on the board, a property that could be exploited in future research. 

\section{Mitigation Techniques}
Despite not being stated as an explicit means of defence, results from \cite{7} imply that simple typing style changes could be sufficient to avoid attack. When touch typing was used, \cite{7} saw keystroke recognition reduce from $64\%$ to $40\%$, which (while still an impressive feat) may not be a high enough accuracy to account for a complex input featuring the shift key, backspace and other non-alphanumeric keys. Additionally, a change in typing style may be implemented alongside mitigation techniques presented in other papers and requires no software or hardware component.

The second simple defence against such attacks would be the use of randomised passwords featuring multiple cases. With the success of language-based models in \cite{3,1,4}, passwords containing full words may be at greater risk of attack. Also, while multiple methods succeeded in recognising a press of the shift key, no paper in the surveyed literature succeeded in recognising the `release peak' of the shift key amidst the sounds of other keys, doubling the search space of potential characters following a press of the shift key.

As stated in section 2.5, the authors of \cite{10} and \cite{16} present methods and therefore countermeasures based on Skype calling. \cite{10} implements two sound-based countermeasures: playing sounds over a speaker near the broadcasting microphone and mixing sounds into the transmitted audio locally. Of the two, the second is more discrete and less distracting for the user. Two types of sound were tested, white noise and fake keystrokes, with the latter proving to be more effective thanks to the sophistication of white noise removal algorithms. The authors in \cite{16} attempted to disrupt keystroke acoustic features by randomly warping the sound slightly whenever keystrokes were detected, a method which reduced accuracy using FFT features to a random guess, but only slightly inhibited MFCC features.

Among the mitigation techniques for voice call attacks, adding randomly generated fake keystrokes to the transmitted audio appears to have the best performance and least annoyance to the user. However, such an approach must only be deployed when keystrokes are detected by the VoIP software as constant false keystrokes may inhibit usability of the software for the receiver. 
One potential direction of future research is the automatic suppression or removal of keystroke acoustics from VoIP applications. Such an implementation would not only defend against ASCAs, but would remove irritating keystroke sounds for the users.

In \cite{3}, the authors recommend a defence which has proven apt with the progression of time in the form of two-factor authentication: utilising a secondary device or biometric check to allow access to data. As more laptops begin to come with biometric scanners built in as standard, the requirement for input of passwords via keyboard is all but eliminated, making ASCAs far less dangerous. However, as stated in \cite{3}, a threat remains that data other than passwords may be retrieved via ASCA. 

Perhaps equally as interesting as effective countermeasures are those presented in papers that have lost viability over time. For example, \cite{2} states touchscreen keyboards present a silent alternative to keyboards and therefore negate ASCAs, however in recent studies compromised smartphone microphones have repeatedly inferred text typed on touchscreens with concerning accuracy \cite{13,14,17}. Similarly, It is recommended in \cite{3} to check a room for microphones before typing private information. Such a technique is nearly entirely negated by the modern ubiquity of microphones. Such a method would require removal of smartphones, smartwatches, laptops, webcams, smart speakers and many more devices from the vicinity. It is stated in \cite{16} that muting their microphone or not typing at all when on a Skype call may defend victims from ASCAs. Such an approach lost some feasibility during the COVID-19 pandemic, at which time a large number of companies began to switch to remote working via video-conference software, necessarily including typing. The diminishing of these countermeasures creates concern that as the prevalence of technology required for these attacks increases, further countermeasures will prove insufficient.

\section{Conclusion}

In this paper, we survey the literature available on acoustic side channel attacks on keyboards. We propose a novel deep learning-based method to perform an effective attack on a keyboard. We collect our data in two modes; by recording via a phone physically co-located with the laptop, and by recording via an online video conferencing tool (Zoom) for a remote attack.

The method presented in this paper achieved a top-1 classification accuracy of $95\%$ on phone-recorded laptop keystrokes, representing improved results for classifiers not utilising language models and the second best accuracy seen across all surveyed literature. When implemented on the Zoom-recorded data, the method resulted in $93\%$ accuracy, an improved result for classifiers using such applications as attack vectors. 

Possible directions of future research include: more robust methods of isolating keystrokes from a single recording, as all ASCA methods rely on accurately isolated keystrokes in order to classify; the use of smart speakers to record keystrokes for classification, as these devices remain always-on and are present in many homes; the implementation of a language model in addition to the method presented in this paper; which could improve keystroke recognition when identifying defined words as well as an end-to-end real-world implementation of an ASC attack on a keyboard.

\section*{Acknowledgement}
This project had full ethical approval from the ethics committee of Durham University, UK. The collected datasets are publicly available for others to perform further research. The attack code is available upon request to researchers. This work has been supported by the PETRAS National Centre of Excellence for IoT Systems Cybersecurity, which has been funded by the UK EPSRC under grant number EP/S035362/1.


%

\bibliographystyle{plain}
\bibliography{references}

\begin{thebibliography}{10}

\bibitem{3GPP}
3GPP~TS 35.205(V4.0.0).
\newblock 3rd generation partnership project; technical specification group services and system aspects;3g security;specification of the milenage algorithm set.
\newblock 2001.

\bibitem{abdulgadir2022side}
Abubakr Abdulgadir, Richard Haeussler, Sammy Lin, Jens-Peter Kaps, and Kris Gaj.
\newblock Side-channel resistant implementations of three finalists of the nist lightweight cryptography standardization process: Elephant, tinyjambu, and xoodyak.
\newblock 2022.

\bibitem{10}
S~Abhishek Anand and Nitesh Saxena.
\newblock Keyboard emanations in remote voice calls: Password leakage and noise (less) masking defenses.
\newblock In {\em Proceedings of the Eighth ACM Conference on Data and Application Security and Privacy}, pages 103--110, 2018.

\bibitem{2}
Dmitri Asonov and Rakesh Agrawal.
\newblock Keyboard acoustic emanations.
\newblock In {\em IEEE Symposium on Security and Privacy, 2004. Proceedings. 2004}, pages 3--11. IEEE, 2004.

\bibitem{1}
Michael Backes, Markus D{\"u}rmuth, Sebastian Gerling, Manfred Pinkal, Caroline Sporleder, et~al.
\newblock Acoustic $\{$Side-Channel$\}$ attacks on printers.
\newblock In {\em 19th USENIX Security Symposium (USENIX Security 10)}, 2010.

\bibitem{11}
Jia-Xuan Bai, Bin Liu, and Luchuan Song.
\newblock I know your keyboard input: A robust keystroke eavesdropper based-on acoustic signals.
\newblock In {\em Proceedings of the 29th ACM International Conference on Multimedia}, pages 1239--1247, 2021.

\bibitem{4}
Yigael Berger, Avishai Wool, and Arie Yeredor.
\newblock Dictionary attacks using keyboard acoustic emanations.
\newblock In {\em Proceedings of the 13th ACM conference on Computer and communications security}, pages 245--254, 2006.

\bibitem{16}
Alberto Compagno, Mauro Conti, Daniele Lain, and Gene Tsudik.
\newblock Don't skype \& type! acoustic eavesdropping in voice-over-ip.
\newblock In {\em Proceedings of the 2017 ACM on Asia Conference on Computer and Communications Security}, pages 703--715, 2017.

\bibitem{29}
Zihang Dai, Hanxiao Liu, Quoc Le, and Mingxing Tan.
\newblock Coatnet: Marrying convolution and attention for all data sizes.
\newblock {\em Advances in Neural Information Processing Systems}, 34, 2021.

\bibitem{38}
Ketan Doshi.
\newblock Audio deep learning made simple: Sound classification, step-by-step, 5 2021.

\bibitem{18}
Jeffrey Friedman.
\newblock Tempest: A signal problem.
\newblock {\em NSA Cryptologic Spectrum}, 35:76, 1972.

\bibitem{genkin2014rsa}
Daniel Genkin, Adi Shamir, and Eran Tromer.
\newblock Rsa key extraction via low-bandwidth acoustic cryptanalysis.
\newblock In {\em Annual cryptology conference}, pages 444--461. Springer, 2014.

\bibitem{gilbert2011testing}
Benjamin~Jun Gilbert~Goodwill, Josh Jaffe, Pankaj Rohatgi, et~al.
\newblock A testing methodology for side-channel resistance validation.
\newblock In {\em NIST non-invasive attack testing workshop}, volume~7, pages 115--136, 2011.

\bibitem{36}
Ian Goodfellow, Jean Pouget-Abadie, Mehdi Mirza, Bing Xu, David Warde-Farley, Sherjil Ozair, Aaron Courville, and Yoshua Bengio.
\newblock Generative adversarial nets.
\newblock {\em Advances in neural information processing systems}, 27, 2014.

\bibitem{15}
Tzipora Halevi and Nitesh Saxena.
\newblock A closer look at keyboard acoustic emanations: random passwords, typing styles and decoding techniques.
\newblock In {\em Proceedings of the 7th ACM Symposium on Information, Computer and Communications Security}, pages 89--90, 2012.

\bibitem{7}
Tzipora Halevi and Nitesh Saxena.
\newblock Keyboard acoustic side channel attacks: exploring realistic and security-sensitive scenarios.
\newblock {\em International Journal of Information Security}, 14(5):443--456, 2015.

\bibitem{33}
Paul Kocher, Joshua Jaffe, and Benjamin Jun.
\newblock Differential power analysis.
\newblock In {\em Annual international cryptology conference}, pages 388--397. Springer, 1999.

\bibitem{35}
Alex Krizhevsky, Ilya Sutskever, and Geoffrey~E Hinton.
\newblock Imagenet classification with deep convolutional neural networks.
\newblock {\em Advances in neural information processing systems}, 25, 2012.

\bibitem{17}
Li~Lu, Jiadi Yu, Yingying Chen, Yanmin Zhu, Xiangyu Xu, Guangtao Xue, and Minglu Li.
\newblock Keylistener: Inferring keystrokes on qwerty keyboard of touch screen through acoustic signals.
\newblock In {\em IEEE INFOCOM 2019-IEEE Conference on Computer Communications}, pages 775--783. IEEE, 2019.

\bibitem{12}
Anindya Maiti, Oscar Armbruster, Murtuza Jadliwala, and Jibo He.
\newblock Smartwatch-based keystroke inference attacks and context-aware protection mechanisms.
\newblock In {\em Proceedings of the 11th ACM on Asia Conference on Computer and Communications Security}, pages 795--806, 2016.

\bibitem{mehrnezhad2015touchsignatures}
Maryam Mehrnezhad, Ehsan Toreini, Siamak~F Shahandashti, and Feng Hao.
\newblock Touchsignatures: Identification of user touch actions based on mobile sensors via javascript.
\newblock In {\em Proceedings of the 10th ACM Symposium on Information, Computer and Communications Security}, pages 673--673, 2015.

\bibitem{mehrnezhad2016touchsignatures}
Maryam Mehrnezhad, Ehsan Toreini, Siamak~F Shahandashti, and Feng Hao.
\newblock Touchsignatures: identification of user touch actions and pins based on mobile sensor data via javascript.
\newblock {\em Journal of Information Security and Applications}, 26:23--38, 2016.

\bibitem{mehrnezhad2018stealing}
Maryam Mehrnezhad, Ehsan Toreini, Siamak~F Shahandashti, and Feng Hao.
\newblock Stealing pins via mobile sensors: actual risk versus user perception.
\newblock {\em International Journal of Information Security}, 17(3):291--313, 2018.

\bibitem{6}
NSA NACSIM.
\newblock 5000: Tempest fundamentals.
\newblock {\em National Security Agency}, 1982.

\bibitem{40}
Daniel~S. Park, William Chan, Yu~Zhang, Chung-Cheng Chiu, Barret Zoph, Ekin~D. Cubuk, and Quoc~V. Le.
\newblock {SpecAugment}: A simple data augmentation method for automatic speech recognition.
\newblock In {\em Interspeech 2019}. {ISCA}, 9 2019.

\bibitem{34}
Adam Paszke, Sam Gross, Francisco Massa, Adam Lerer, James Bradbury, Gregory Chanan, Trevor Killeen, Zeming Lin, Natalia Gimelshein, Luca Antiga, Alban Desmaison, Andreas Kopf, Edward Yang, Zachary DeVito, Martin Raison, Alykhan Tejani, Sasank Chilamkurthy, Benoit Steiner, Lu~Fang, Junjie Bai, and Soumith Chintala.
\newblock Pytorch: An imperative style, high-performance deep learning library.
\newblock In H.~Wallach, H.~Larochelle, A.~Beygelzimer, F.~d\textquotesingle Alch\'{e}-Buc, E.~Fox, and R.~Garnett, editors, {\em Advances in Neural Information Processing Systems 32}, pages 8024--8035. Curran Associates, Inc., 2019.

\bibitem{27}
Joseph Roth, Xiaoming Liu, Arun Ross, and Dimitris Metaxas.
\newblock Investigating the discriminative power of keystroke sound.
\newblock {\em IEEE Transactions on Information Forensics and Security}, 10(2):333--345, 2014.

\bibitem{39}
Connor Shorten and Taghi~M Khoshgoftaar.
\newblock A survey on image data augmentation for deep learning.
\newblock {\em Journal of big data}, 6(1):1--48, 2019.

\bibitem{14}
Ilia Shumailov, Laurent Simon, Jeff Yan, and Ross Anderson.
\newblock Hearing your touch: A new acoustic side channel on smartphones.
\newblock {\em arXiv preprint arXiv:1903.11137}, 2019.

\bibitem{24}
Fran{\c{c}}ois-Xavier Standaert.
\newblock Introduction to side-channel attacks.
\newblock In {\em Secure integrated circuits and systems}, pages 27--42. Springer, 2010.

\bibitem{13}
Kai~Ren Teo, BT~Balamurali, Chen~Jer Ming, and Jianying Zhou.
\newblock Retrieving input from touch interfaces via acoustic emanations.
\newblock In {\em 2021 IEEE Conference on Dependable and Secure Computing (DSC)}, pages 1--8. IEEE, 2021.

\bibitem{30}
Ehsan Toreini, Brian Randell, and Feng Hao.
\newblock An acoustic side channel attack on enigma.
\newblock {\em School of Computing Science Technical Report Series}, 2015.

\bibitem{37}
Ashish Vaswani, Noam Shazeer, Niki Parmar, Jakob Uszkoreit, Llion Jones, Aidan~N Gomez, {\L}ukasz Kaiser, and Illia Polosukhin.
\newblock Attention is all you need.
\newblock {\em Advances in neural information processing systems}, 30, 2017.

\bibitem{31}
Martin Vuagnoux and Sylvain Pasini.
\newblock Compromising electromagnetic emanations of wired and wireless keyboards.
\newblock In {\em USENIX security symposium}, volume~8, pages 1--16, 2009.

\bibitem{5}
Peter Wright.
\newblock Spycatcher: The candid autobiography of a senior intelligence officer.
\newblock {\em New York: Viking}, 1987.

\bibitem{41}
Xiaoma Xmu.
\newblock External-attention-pytorch/coatnet.py at master · xmu-xiaoma666/external-attention-pytorch, 10 2021.

\bibitem{32}
Yuval Yarom and Katrina Falkner.
\newblock $\{$FLUSH+ RELOAD$\}$: A high resolution, low noise, l3 cache $\{$Side-Channel$\}$ attack.
\newblock In {\em 23rd USENIX security symposium (USENIX security 14)}, pages 719--732, 2014.

\bibitem{9}
Tong Zhu, Qiang Ma, Shanfeng Zhang, and Yunhao Liu.
\newblock Context-free attacks using keyboard acoustic emanations.
\newblock In {\em Proceedings of the 2014 ACM SIGSAC conference on computer and communications security}, pages 453--464, 2014.

\bibitem{3}
Li~Zhuang, Feng Zhou, and J~Doug Tygar.
\newblock Keyboard acoustic emanations revisited.
\newblock {\em ACM Transactions on Information and System Security (TISSEC)}, 13(1):1--26, 2009.

\end{thebibliography}

\end{document}